\documentclass[aip,rsi,reprint,graphicx]{revtex4-1} 

\usepackage{amsmath, newtxtext, newtxmath}

\usepackage[pdftex]{graphicx}
\usepackage{dcolumn}
\usepackage{bm}
\usepackage{soul}

\usepackage{bm}
\usepackage{soul}

\usepackage{hyperref}
\hypersetup{
    colorlinks=true,        
    linkcolor=blue,          
    citecolor=blue,        
    filecolor=magenta,      
    urlcolor=blue           
}
\usepackage{color}
\newcommand{\red}{\textcolor{black}}
\newcommand{\blue}{\textcolor{black}}

\draft 

\begin{document}


\title{Note: Optical filter method for high-resolution magnetostriction measurement using fiber Bragg grating under millisecond-pulsed high magnetic fields at cryogenic temperatures} 



\author{Akihiko Ikeda}
\email[e-mail: ]{ikeda@issp.u-tokyo.ac.jp}
\author{Yasuhiro H. Matsuda}
\email[e-mail: ]{ymatsuda@issp.u-tokyo.ac.jp}

\affiliation{Institute for Solid State Physics (ISSP), University of Tokyo (UTokyo), Kashiwa, Japan}
\author{Hiroshi Tsuda}
\affiliation{National Institute of Advanced Industrial Science and Technology, Tsukuba, Japan}

\date{\today}

\begin{abstract}

High-resolution magnetostriction measurement of $\Delta L/L\sim10^{-6}$ at \red{a} speed of 5 MHz \red{is performed,} using optical filter method as the detection scheme for the fiber Bragg grating (FBG) based strain monitor is performed under 35-millisecond pulsed high magnetic fields up to 45 T at 2.2 K.
The resolution of magnetostriction is about the same order as the conventionally reported \red{value from} FBG based magnetostriction measurement systems for millisecond pulsed magnetic fields.
The measurement speed is $\sim$100 times the conventional ones.
Present system can be a \blue{faster alternative for} the conventional FBG based magnetostriction measurement system for millisecond pulsed high magnetic fields.

\end{abstract}

\pacs{}

\maketitle 


Various materials have been found to undergo interesting phase transitions \red{in presence of high magnetic fields, in most cases, accompanied by modification} of their crystal structure due to spin-lattice coupling.
Examples of \red{such modifications} are the exchange striction of magnetic insulators, volume expansion by valence or spin state transition, magnetic anisotropy through spin-orbit coupling, and so forth.
\red{Therefore, for gaining insights into the spin-lattice coupling itself and to detect novel phase changes, magnetostriction measurement at high magnetic fields is indispensable.}

\blue{
Magnetostriction is measured using variety of methods.
Resistive strain gauge, capacitance dilatometory,\cite{Kuchler} scanning tunneling microscopy \cite{costa} and fiber Bragg grating (FBG) \cite{jaimesensors} measures magnetostriction with a resolution of $\Delta L/L < 10^{-8}$ under static magnetic fields.
However, under pulsed high magnetic fields, the resolution generally gets poorer ($\Delta L/L > 10^{-6}$) due to the limited time and space, and noises from mechanical and electrical origin.\cite{Doerr, daou, jaimesensors}
}

\begin{figure}
\begin{center}
\includegraphics[scale=0.45]{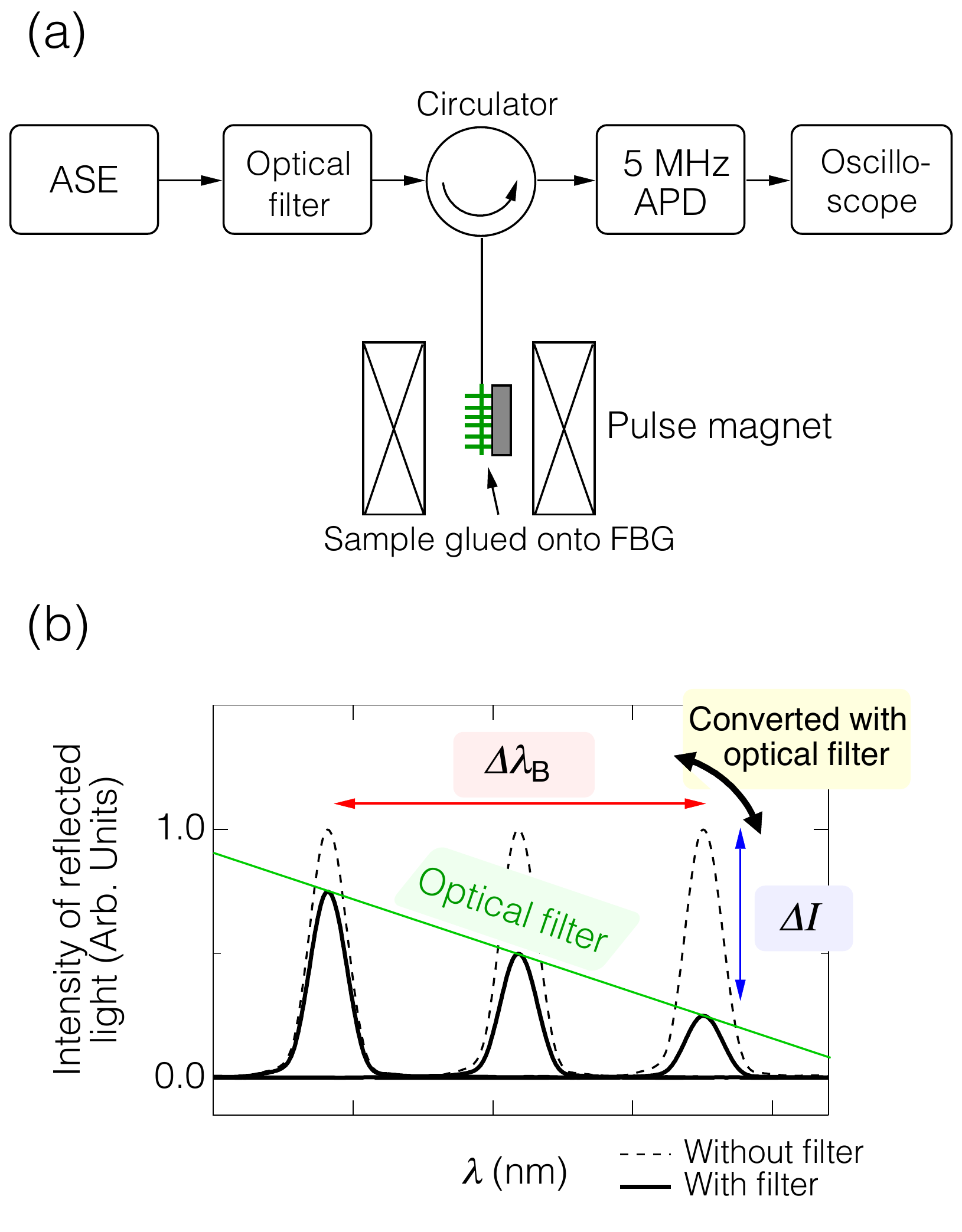}
\caption{
(a) Schematic drawing of the devised FBG based strain measurement system in the high-resolution mode.
\red{Light from t}he broad band light source is filtered with \red{an} optical filter.
The filtered light illuminates the FBG whose reflection is detected with an avalanche photodiode.
(b) \red{A spectrum illustrating output of the} optical filter method for the FBG based strain measurement.
The change of Bragg wavelength of the FBG $\Delta \lambda_{\rm{B}}$ is converted to $\Delta I$ using the optical filter.
The change of the intensity of the reflection originates in the shift of \red{wavelength,} $\Delta \lambda_{\rm{B}}$.\label{sche}}
\end{center}
\end{figure}

FBG is an optical single mode fiber with a Bragg grating at its core with axial length of a few millimeters, enabling us to detect \red{the fine strain $\Delta L/L$ of material} glued to the FBG fiber.
\red{This is achieved} by detecting the change \red{in} the Bragg wavelength \red{, which is directly proportional to the strain as given by the relation} $\Delta \lambda_{\rm{B}}/\lambda_{\rm{B}} \propto \Delta L/L$.
FBG based magnetostriction measurement has been reported to be extremely feasible with pulsed high magnetic fields,\cite{daou, jaimepnas, rotter, jaimesensors} owing to the following facts, (a) it is immune to electric noise, (b) the high-quality of the commercially available products, (c) it consumes only small volume and small samples are measurable, and (d) absolute value of strain is obtained \red{directly} without calibration.

Detection of $\Delta \lambda_{\rm{B}}/\lambda_{\rm{B}}$ \red{depends} on high-speed InGaAs spectrometers with broadband light sources in the conventionally reported magnetostriction systems for millisecond-pulsed high magnetic fields.\cite{daou, jaimesensors}
The resolution and the speed of the measurement is limited by the specification of the detectors, which is reported to be $\Delta L/L <1\times10^{-6}$ at $\sim50$ kHz.
On the other hand, optical filter method has been reported to realize high-speed response of FBG based strain monitor as an alternative method to the spectroscopic method. \cite{tsuda}
In optical filter method, as schematically shown in Fig. \ref{sche}(a) and \ref{sche}(b), the shift of the FBG peak position $\Delta \lambda_{\rm{B}}$ is converted to the intensity of the reflected light $\Delta I$, which is detected with \red{photoreceivers}.
\red{We have earlier} devised a high-speed 100 MHz strain monitor with the optical filter method with the resolution of $\Delta L/L\sim1\times10^{-4}$.\cite{ikedafbg}
We showed its feasibility under ultrahigh magnetic fields of up to 150 T that can be generated only with destructive pulse magnets whose time duration is limited to a few microseconds. \cite{herlach}

In this note, we show a high resolution magnetostriction measurement of $\Delta L/L\sim1\times10^{-6}$ at a speed of 5 MHz using the optical filter method for the FBG based magnetostriction measurement system.
The method is shown to be quite useful with millisecond-pulsed high magnetic fields.
As a working example, a magnetostriction measurement of volborthite, a \red{two-dimensional (2D)} quantum magnet, is shown at 2.2 K \red{in fields} up to 45 T, which is compared with the magnetization.
\blue{Thus, the optical filter method is shown to be a faster alternative for the conventional detection scheme for the FBG based magnetostriction measurement system for millisecond pulsed high magnetic fields.}

\begin{figure}
\begin{center}
\includegraphics[scale=0.55]{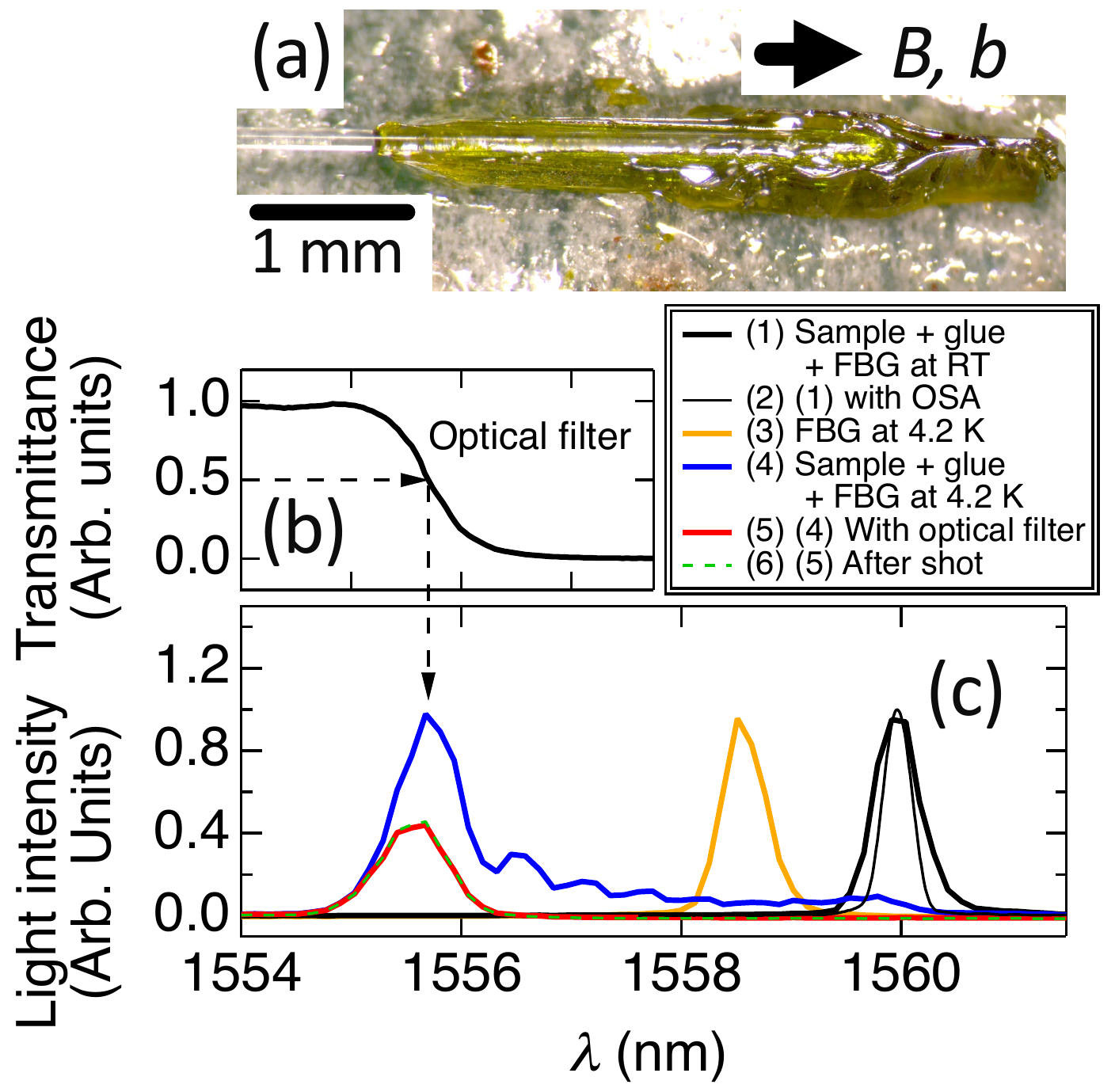}
\caption{
\blue{
(a) A photo of a single crystalline sample of volborthite on top of which an optical fiber with FBG is glued using acrylic type glue.
The axis of the optical fiber, external magnetic field, and the crystallographic  $b$ axis of volborthite orient in the same direction as shown by the arrow \red{in the picture}.
(b) Transmittance spectrum of the optical filter used in the optical filter method.
(c) Spectra of reflected light from FBG under various conditions as is indicated in the legend.
All spectra are measured with an infrared spectrometer except for spectrum (2) which is measured using an optical spectrum analyzer (OSA) shown for comparison.
}
\label{spectra}
}
\end{center}
\end{figure}

The schematic diagram of a FBG based magnetostriction measurement system using optical filter method is shown in Fig. \ref{sche}(a).
\red{Light from t}he broadband light source is the amplified stimulated emission (ASE) light source (Amonics C-Band ASE Light Source ALS-15-B-FA).
The light source is filtered with a sharp optical filter (Koshin Kogaku 1560SPF) installed in a tunable filter module (Koshin Kogaku TFM/FC).
The filtered light illuminates the FBG on which the sample is glued.
The reflection from the FBG is guided with an optical circulator and detected with an avalanche photodiode (APD) (Thorlab PDA20C/M) at 5 MHz.

To demonstrate the resolution of the magnetostriction measured with the developed FBG based measurement in optical filter method, longitudinal magnetostriction of volborthite, Cu$_{3}$V$_{2}$O$_{7}$(OH)$_{2}\cdot$2H$_{2}$O, is measured.
Note that the linearity \red{and} precision of strain measured using FBG \cite{daou} in the optical filter method have already been confirmed in earlier studies.\cite{tsuda}
Volborthite is a geometrically frustrated 2D quantum spin system on a deformed Kagom\'{e} lattice.\cite{ishikawa}
We \red{chose} this sample because the magnetostriction of inorganic quantum spin system is significant and informative.
Firstly, it is expected to be \red{so} small that almost no studies on magnetostriction of \red{this} quantum spin system has been reported in the range of $\Delta L/L\sim10^{-6}$.
Secondly, it is directly connected with its spin correlation function which is \red{inaccessible} with other methods at such high magnetic fields.\cite{vivian}
Pulsed high-field facility in ISSP, UTokyo is used for magntic field generations.

In the case of volborthite, the bare optical fiber with FBG is glued to a single crystalline sample using an acrylic type glue.
Sample geometry is 4$\times$0.4$\times$0.2 mm$^{3}$ and the glass fiber with FBG has a diameter of 120 $\mu$m as shown in Fig. \ref{spectra}(a).
The magnetostriction measurement is successful with this gluing condition.
In general, valid glueing conditions may vary between samples, \red{sample dimensions,} and surface conditions.
Epoxy type glues work best in our experience. 
However, acrylic type glues is also useful in the sense that it is removable and that the FBG and the sample \red{are} reusable after the experiment unlike when using the epoxy type glues.
Our current strategy is that we first try to use the acrylic type glue, cool it down to liquid He temperature\red{, and then apply} the pulsed magnetic field.
If it fails somewhere in the process, we try with a stronger epoxy glue in the next time.

\begin{figure}
\begin{center}
\includegraphics[scale=0.53]{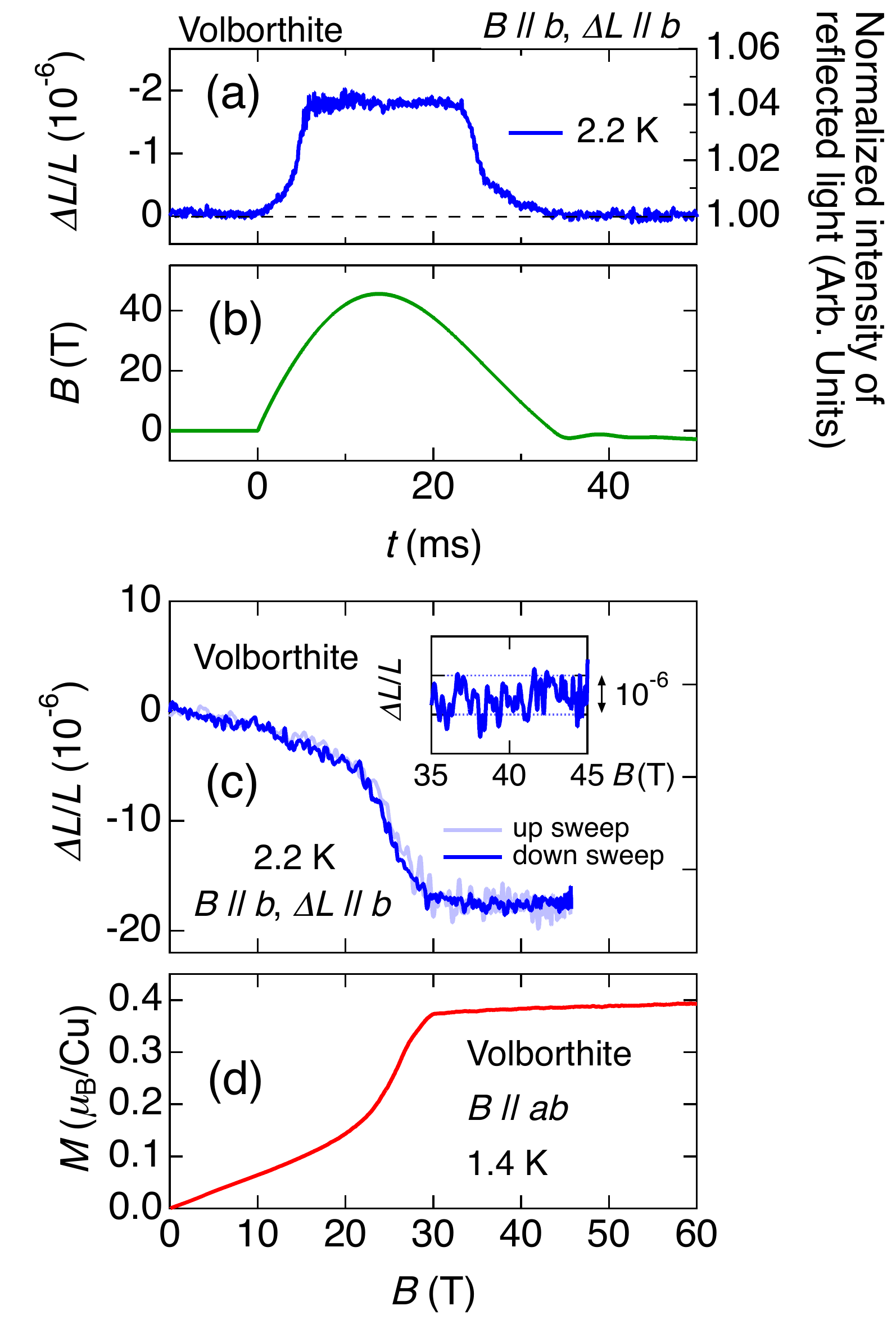}
\caption{
\blue{
(a) Time evolution of the intensity change of the reflected light from the FBG.
(b) The waveform of a pulsed high magnetic field with a 35 ms time duration. 
(c) Longitudinal magnetostriction of volborthite deduced using the data in (a).
Inset shows the magnification of the magnetostriction at the 1/3 magnetization plateau region. 
(d) The magnetization curve of volborthite. \cite{yoshida}
}
\label{volbo}}
\end{center}
\end{figure}

\blue{In Figs. \ref{spectra}(b) and \ref{spectra}(c), transmittance spectrum of the optical filter and spectra of the reflection from the FBG are shown, respectively.}
The FBG used has a resonance peak at 1560 nm at room temperature.
It remains the same after the FBG is glued to a sample of volborthite.
It shows a blue shift after being cooled down to 4.2 K.
Note that the yield of the blue shift is significantly larger for FBG with the sample as shown in the spectrum (4) in \blue{Fig. \ref{spectra}(c)} compared \red{to} that without the sample as shown in the spectrum (3) in \blue{Fig. \ref{spectra}(c)}.
The tailing structure of the spectrum shown in the spectrum (4) \blue{Fig. \ref{spectra}(c)} should originate \red{due to} the fact that the FBG is a bit longer ($\sim4$ mm) than the sample ($\sim2.5$ mm) in the present case.
The tailing structure does not affect the measurement because it is eliminated by the optical filter inserted next to the light source \red{as is clear from} \blue{Fig. \ref{spectra}(c)} whose transmittance spectrum is shown in \blue{Fig. \ref{spectra}(b)}.
The central position of the optical filter is adjustable with the tunable optical filter module and \red{it is positioned such} that the FBG peak position is located at the transmittance of 0.5.
With that configuration, one is able to detect elongation and contraction of the sample \red{as the intensity of the reflected light decreased and increased}, respectively.

In Fig. \ref{volbo}, results of magnetostriction measurement of volborthite at 2.2 K are shown.
\blue{Figure \ref{volbo}(a) and \ref{volbo}(b) show the time evolution of the intensity of the reflected light from the FBG and the pulsed magnetic field of up to 45 T with 35 ms time duration, respectively, the former of which is proportional to magnetstriction in the present method as shown in Fig. \ref{sche}(b).}
The intensity of reflected light increases with increasing magnetic field and becomes static \red{for} magnetic fields larger than $\sim30$ T.
The increase of the intensity of the reflected light indicates that the sample shrinks under magnetic fields within the present configuration of the optical filter as shown in Fig. \ref{sche}(b).
The intensity of reflected light decreases \red{on} decreasing \red{the} magnetic field and returns to the original intensity after the external magnetic field is removed.

\blue{In Fig. \ref{volbo}(c)}, longitudinal magnetostriction of volborthite at 2.2 K up to 45 T is shown, which is deduced \red{using} the data in \blue{Figs. \ref{volbo}(a) and \ref{volbo}(b)}.
The absolute value of magnetostriction is deduced by a simple form, $\Delta L/L=c_{1}\Delta \lambda_{\rm{B}}/\lambda_{\rm{B}}$ and $\Delta \lambda_{\rm{B}}/\lambda_{\rm{B}}=c_{2} \Delta I_{\rm{R}}/I_{\rm{R}}$, where $c_{1}$ is a strain-optic collection term,\cite{daou} $I_{\rm{R}}$ is the intensity of the reflected light and $c_{2}$ is the conversion ratio of the intensity and $\lambda_{\rm{B}}$ of the FBG.
Values of $c_{1}=1.316$\cite{daou} and $c_{2}=1/3000$ \red{are} used in the present study.
The conversion ratio of the optical filter $c_{2}$ can be varied by using different optical filters, resulting in the variation of  the resolution and the dynamic range of measurable $\Delta L/L$.\cite{ikedafbg, ikedafbg2}

The obtained $\Delta L/L$ of volborthite \blue{shown in Fig. \ref{volbo}(c)} is compared with the magnetization curve at 1.4 K \blue{as shown in Fig. \ref{volbo}(d).} \cite{yoshida}
$\Delta L/L$ shows a gentle decrease in the region of 0 T to 22 T, a steep decrease in the region of 22 T to 29 T and a plateau above 29 T.
The resultant resolution of the measurement is $\sim1\times10^{-6}$ at 5 MHz as shown in the inset in Fig. \ref{volbo}(c).
\blue{On the other hand,} the magnetization of volborthite starts to increase from 0 T with increasing magnetic field.
A relatively steep increase is observed from $\sim22$ T to $\sim29$ T and it enters a 1/3 magnetization plateau at $\sim29$ T which continues at least up to $\sim75$ T.\cite{ishikawa}
We \red{find that} the trend of the magnetization curve is in good agreement with the magnetostriction observed in the present study.
The $\Delta L/L$ \red{variation is in} even better agreement with the $M^{1.3}$ curve (not shown).
The contraction of the lattice with increasing magnetization is considered to originate in the decreasing Cu-O-Cu bond angle that enhances the ferromagnetic superexchange interaction as also discussed in the case of SrCu(BO$_{3}$)$_{2}$. \cite{radtke}
Detailed discussion of the results will be reported elsewhere.

In summary, we show a high resolution magnetostriction measurement  of $\Delta L/L\sim1\times10^{-6}$ at 5 MHz with the FBG based magnetostriction measurement in optical filter method which operates under millisecond pulsed high magnetic fields.
Magnetostriction measurement of a frustrated 2D quantum magnet, volborthite, up to 45 T at 2.2 K is presented as a working example.
\blue{Thus, the optical filter method is shown to be a faster alternative for the conventional detection scheme for the FBG based magnetostriction measurement system for millisecond pulsed high magnetic fields.}

We thank H. Ishikawa of Max Planck Institute for Solid State Research and Z. Hiroi of ISSP, UTokyo for providing the sample of volborthite.
T. Yajima of the X-ray facility in ISSP, UTokyo is acknowledged for its help in determining the crystalline axes.
We thank S. Takeyama of ISSP, UTokyo for experimental and other support.
This work was supported by JSPS KAKENHI Grant-in-Aid for Young Scientists (B) Grant No. 16K17738, Grant-in-Aid for Scientific Research (B) Grant No. 16H04009 and the internal research grant from ISSP, UTokyo. 

\bibliography{fbg}

\end{document}